# Highly tunable quadruple quantum dot in a narrow bandgap semiconductor InAs nanowire

Jingwei Mu,[a,c] Shaoyun Huang,[a] Zhi-Hai Liu,[a] Weijie Li,[a,c] Ji-Yin Wang,[a] Dong Pan,[b,d] Guang-Yao Huang,[a] Yuanjie Chen,[a] Jianhua Zhao,[b,d,†] and H. Q. Xu[a,c,d,*]

[a] *Beijing Key Laboratory of Quantum Devices, Key Laboratory for the Physics and Chemistry of Nanodevices and Department of Electronics, Peking University, Beijing 100871, China*

[b] *State Key Laboratory of Superlattices and Microstructures, Institute of Semiconductors, Chinese Academy of Sciences, P.O. Box 912, Beijing 100083, China*

[c] *Academy for Advanced Interdisciplinary Studies, Peking University, Beijing 100871, China*

[d] *Beijing Academy of Quantum Information Sciences, Beijing 100193, China*

Corresponding authors. Emails: *hqxu@pku.edu.cn; †jhzhao@semi.ac.cn

(January 18, 2020)

## ABSTRACT

Quantum dots (QDs) made from semiconductors are among the most promising platforms for the developments of quantum computing and simulation chips, and have advantages over other platforms in high density integration and in compatibility to the standard semiconductor chip fabrication technology. However, development of a highly tunable semiconductor multiple QD system still remains as a major challenge. Here, we demonstrate realization of a highly tunable linear quadruple QD (QQD) in a narrow bandgap semiconductor InAs nanowire with fine finger gate technique. The QQD is studied by electron transport measurements in the linear response regime. Characteristic two-dimensional charge stability diagrams containing four groups of resonant current lines of different slopes are found for the QQD. It is shown that these current lines can be individually assigned as arising from resonant electron transport through the energy levels of different QDs. Benefited from the excellent gate tunability, we also demonstrate tuning of the QQD to regimes where the energy levels of two QDs, three QDs and all the four QDs are energetically on resonance, respectively, with the fermi level of source and drain contacts. A capacitance network model is developed for the linear QQD and the simulated charge stability diagrams based on the model show good agreements with the experiments. Our work presents a solid experimental evidence that narrow bandgap semiconductor nanowires multiple QDs could be used as a versatile platform to achieve integrated qubits for quantum computing and to perform quantum simulations for complex many-body systems.

## 1. Introduction

Quantum dots (QDs) made from semiconductors are among the most promising building blocks for the developments of solid-state based quantum computing[1-3] and quantum simulation technologies.[4-6] In comparison with other quantum-chip technology platforms, semiconductor QD-based architectures have advantages in achieving high-density qubit integration and in compatibility with the state of the art chip fabrication technology. In a semiconductor QD system, the spin and/or charge degrees of freedom of localized electrons are employed in realization of qubits[1-3] and in performing quantum simulation.[4-6] Extensive experimental efforts have been placed on the fabrication and characterization of semiconductor double QDs (DQDs) and triple QDs (TQDs),[7-12] and on coherent manipulations of spin qubits[13-18] or charge qubits[2,19,20] made from them. Realization of a scalable multiple QD system is then becoming one of the major challenges for further development of integrated quantum chips. Scaling up to a quadruple QD (QQD) is a natural key step toward this direction.

Very recently, QQDs made either in a square[21,22] or a linear[23-27] configuration in two-dimensional electron systems (2DESs) have been realized and studied. However, QQDs made from narrow bandgap semiconductor nanowires have been seldom addressed, although QDs built from these narrow bandgap semiconductor nanowires possess large Landé g-factors and strong spin-orbit interaction,[28-31] allowing for rapid manipulations of spin qubit states by all-electrical means.[32-34] Coherent manipulations of single spin qubits have been demonstrated in InAs and InSb nanowire double QDs.[33,34] A QQD in an InAs nanowire will provide a promising platform for the realization of two or more spin qubits. In addition, QDs built from semiconductor nanowires by fine finger gate technique also have advantages over their 2DES counterparts in tunability for inter-dot couplings and charge state configurations.[10,12,33-38] Thus, an InAs nanowire QQD could also be used as a pioneer system for studying the fundamental physics of electron-electron interactions such as the Fermi-Hubbard model[4,5,8,39,40] and the negative spin exchange coulping.[41,42]

In this work, we demonstrate realization of a linear QQD in a narrow bandgap semiconductor InAs nanowire via finger gate technique. The charge state configuration and tunability of the linear QQD are studied by electron transport measurements. The measurements show that the charge states in individual QDs can be mapped out from the resonant current lines of different slopes found in measured two-dimensional charge stability diagrams of the QQD. Benefited from the excellent gate tunability, we also demonstrate tuning of the QQD to regimes where the energy levels of two QDs, three QDs and all the four QDs are energetically on resonance with the fermi level of the source and drain contacts. A capacitance network model is developed for the linear QQD and the simulated charge stability diagrams based on the model show good agreements with the experiments.

## 2. Device and methods

Fig. 1 shows a scanning electron microscope (SEM) image and a cross-sectional schematic view of a fabricated InAs nanowire QQD device. The InAs nanowires employed in this work are grown by molecular beam epitaxy.[43] The low-temperature transport measurements in a previous work indicate that these InAs nanowires can be regarded as quasi-one-dimensional systems.[44] For device fabrication, finger gate arrays are prepared on a heavily p-doped Si substrate capped with a 200-nm-thick layer of $SiO_2$. Here, electron-beam lithography (EBL) is used to define the fine patterns of the finger gate arrays on the resist. A metal bilayer of 5-nm-thick titanium and 10-nm-thick gold is then deposited onto the sample by electron-beam evaporation (EBE). After lift-off, the finger gate arrays are obtained. In the present work, each array contains fifteen finger gates with a width of ~30 nm and a pitch of ~70 nm [see Fig. 1(a)]. Subsequently, a 10-nm-thick $HfO_2$ layer is deposited onto the sample via atomic layer deposition and InAs nanowires are then transferred onto the sample by a dry method. The contact areas of nanowires are defined through a second step of EBL. Then, soft oxygen-plasma milling is applied to eliminate resist residues and chemically etching in a diluted $(NH_4)_2S_x$ solution is performed to remove surface oxides of the nanowires. The source and drain electrodes are made by deposition of 5-nm-thick titanium and 90-nm-thick gold via EBE again and the device fabrication is finally completed after lift-off. The obtained devices are first electrically characterized in a probe station at room temperature. The devices with a source-drain resistance of less than 50 kΩ, i.e., with InAs nanowires in good contacts to the source and drain electrodes and at an open state without applying a gate voltage, are selected for cryogenic measurements. In the present work, we focus on the results of our cryogenic measurements for the device made from an InAs nanowire of ~30 nm in diameter as shown in Fig. 1(a).

Our cryogenic transport measurements are performed in a dilution refrigerator at a base temperature of ~25 mK. As shown in Fig. 1(a), nine finger gates labeled as G1 to G9 from left to right are located below the InAs nanowire between source and drain electrodes. A bias voltage is applied to the drain electrode with the source electrode being grounded [see Fig. 1(b)]. Gate G2 does not work efficiently and is therefore kept unused throughout the measurements. The other eight gates (G1 and G3 to G9) are able to work as desired. The transfer characteristics of these eight gates are measured and a current pinch-off voltage is found for each individual gate to be in the range of −3.0 to −4.5 V at source-drain bias voltage $V_{ds}$ = 0.5 mV. Each of these finger gates can be used to define a tunneling barrier in the nanowire by applying a voltage around its pinch-off voltage. A single QD in the InAs nanowire can be created using two finger gates. For example, using finger gates G3 and G5 (see Part I of Supplementary Materials), a QD with an averaged electron addition energy of ~6.5 meV and quantization energy of ~1.5 meV can be defined in the InAs nanowire. Multiple QDs can be constructed by applying an appropriate combination of voltages to a set of selected finger gates to form tunneling barriers in the nanowire.

## 3. Results and discussion

### *3.1. Charge stability diagrams of two DQDs defined in the InAs nanowire*

Figs. 2(a) and 2(b) show the charge stability diagrams of two DQDs, referred as the left and right DQDs, defined in the InAs nanowire as illustrated in the cross-sectional schematics shown in Figs. 2(c) and 2(d). Here, gates G1, G3, G4 and G5 are used to generate the left DQD (contains two single QDs, which we label as QD1 and QD2). As schematically shown in Fig. 2(c), gates G1 and G5 are used to define the two outer tunneling barriers, gate G4 is employed to tune the electrostatic potential of QD2, and gate G3 is used to tune both the inter-dot coupling between the two QDs and the electrostatic potentials of the two QDs.[36,45] Fig. 2(a) shows the source-drain current $I_{ds}$ of the left DQD at $V_{ds}$ = 0.1 mV as a function of voltages $V_{G3}$ and $V_{G4}$ applied to gates G3 and G4. It is seen that resonant current lines form good hexagon-shaped patterns and the DQD is in the weak inter-dot coupling regime. Finite current along the boundaries of the hexagons is caused by co-tunneling processes when the energy levels of any one of the two QDs are on resonance with the Fermi level of the source and drain electrodes.[7] Fig. 2(b) displays the source-drain current $I_{ds}$ of the right DQD (contains two single QDs, which we label as QD3 and QD4), defined using gates G5, G7 and G9, as a function of voltages $V_{G6}$ and $V_{G8}$ applied to gates G6 and G8, at $V_{ds}$ = 0.1 mV. Here, as schematically shown in Fig. 2(d), the two outer tunneling barriers are controlled by gates G5 and G9, the inter-dot tunneling barrier is controlled by gate G7, and the numbers of confined electrons in the two QDs are controlled by gates G6 and G8 (two plunger gates). The transport measurements of the two DQDs demonstrate that the finger gates have an excellent controllability and can readily be used to define a linear tunable QQD.

### *3.2. Charge stability diagrams of a QQD built in the InAs nanowire*

Fig. 3(a) displays the charge stability diagram of a QQD constructed in the InAs nanowire as schematically shown in Fig. 1(b). Here, gates G1 and G9 are used to define the two outermost tunneling barriers of the QQD and control the coupling of the drain electrode to QD1 and the coupling of the source electrode to QD4, respectively. Gates G3, G5 and G7 are used to define the inter-dot barriers within the QQD and control the coupling strengths of neighboring QDs. Additionally, gates G3, G5 and G7 have also played a role in altering the electrostatic potentials of the four QDs in the QQD. Gates G4, G6 and G8 are used as plunger gate to tune the electron numbers in QD2, QD3 and QD4, respectively. Here, we note that the cross-talks between a plunger gate and QDs other than its local QD are comparatively weak in this device (see Part II of Supplementary Materials). Thus, in order to map out the charge state configuration of the QQD in a two-dimensional charge stability diagram effectively, we choose to measure the source-drain current $I_{ds}$ as a function of the voltages $V_{G3}$ and $V_{G7}$ applied to the barrier gates G3 and G7 in this work,[10,36,45] other than the voltages applied to the two plunger gates which are usually scanned in the measurements of a QQD defined in a 2DES.[23,24] The current lines seen in the measured charge stability diagram shown in Fig. 3(a) derive from co-tunneling processes when the energy levels of one of the four QDs are on resonance with the Fermi level of the source and drain electrodes. These resonant current lines can be categorized into four groups (labeled as groups 1 to 4) of different slopes, as marked with the dashed lines of different colors, in the charge stability diagram. A current line in each group corresponds to electron filling in and extracting from the same QD. The slope of the current line is determined by the coupling strengths of the corresponding

QD to gates G3 and G7, which dominantly depend on the geometrical distances between the QD and the two finger gates.[10] Thus, current lines in groups 1 and 2 (3 and 4) can be assigned to be associated with co-tunneling through the resonant states of QD1 and QD2 (QD3 and QD4), which couple strongly to gate G3 (G7) but weakly to gate G7 (G3). We also find that current lines in group 1 (4) are approximately perpendicular to the $V_{G3}$ ($V_{G7}$) axis, implying a very weak coupling of the associated QD to gate G7 (G3). As a consequence, we can firmly determine that current lines in group 1 correspond to resonant electron transport through QD1, see Fig. 3(b) for an energy diagram of the QD levels in this case. Similarly, current lines in group 4 can be assigned with resonant electron transport through QD4. Current lines in group 2 (3) with a slightly larger inclination compared to current lines in group 1 (4) can be assigned to resonant electron transport through QD2 (QD3), which in comparison to QD1 (QD4) is geometrically closer to gate G7 (G3).

The above inferences can be double-checked by charge stability diagram measurements at different individual plunger gate voltages. For example, gate G6, plunger gate of QD3, possesses a strong coupling to QD3, but relatively weak cross-talks to the other QDs. We, therefore, tune $V_{G6}$ (the voltage applied to gate G6), and examine the changes of current lines in the measured charge stability diagrams. Fig. 3(d) shows the right half of the charge stability diagram of Fig. 3(a), i.e., the measurements at $V_{G6} = -1.05$ V, while Figs. 3(e) and 3(f) show the measured charge stability diagrams at $V_{G6} = -1.046$ and $-1.042$ V. It can be seen that the positions of the current lines in groups 1, 2 and 4 remain almost unchanged, while the positions of the current lines in group 3 move toward more negative values of $V_{G7}$ with increasing $V_{G6}$. Thus, the current lines in group 3 evidently arise from resonant electron transport through QD3. Similar verifications of our assignments of the current lines have also been carried out, see Part III of Supplementary Materials. Owing to the weak cross-talks between a plunger gate and QDs other than its local QD, we can also easily achieve a selective accessibility of a QD in the nanowire QQD by precisely tuning its own plunger gate without significantly affecting other QDs. It is important to note that in Fig. 3(d), 3(e) or 3(f), two current lines of group 3, marked by M and N, are observable and current line N is seen to move relatively faster than current line M with increasing $V_{G6}$. This is because the energy level of QD3 represented by current line M is closer to an energy level of QD4 and experiences a stronger level repulsion from the energy level, as compared to the energy level of QD3 represented by current line N.

In the measured charge stability diagram shown in Fig. 3(a), a few specific locations, e.g., those labeled by I, III and IV, are present, at which the energy levels of two QDs are aligned with each other and with the Fermi level of the source and drain electrodes, see the schematic shown in Fig. 3(c) for the case at location I, assuming that the interaction between the two QD levels is negligible. At such a location, a cross of two current lines of different slopes is primarily observed. However, at locations, such as one labeled by II, two current lines of different slopes are seen to show an avoiding crossing. This is due to the fact that the two current lines arise from resonant electron transport through the energy levels of two adjacent QDs (e.g., QD2 and QD3 at location II) and a relatively strong coupling between the two energy levels is present.

### *3.3. Electrostatic capacitance network model for the InAs nanowire QQD*

Fig. 4(a) displays a simplified equivalent circuit diagram for our QQD device, which is described as a network of capacitors and resistors.[7,8,9] The circuit diagram includes source and drain electrodes, four gates (G3, G4, G6 and G7) that are changed in the measurements, and four QDs (QD1, QD2, QD3 and QD4) that act as four nodes and are marked by colored ovals. We take into account three kinds of major mutual capacitances: (1) the capacitances ($C_{G3\text{-}QDi}$ and $C_{G7\text{-}QDi}$) between QDi (i=1, 2, 3 or 4 ) and gates G3 and G7 that are scanned in the measurements, (2) the capacitance $C_{G4\text{-}QD2}$ ($C_{G6\text{-}QD3}$) between gate G4 (G6) and QD2 (QD3) that is adjusted to tune the QQD to different regimes of interest, and (3) the capacitances ($C_{Mij}$) between QDi and QDj (i,j=1, 2, 3 or 4 and j≠i). The Coulomb interaction between the electrons in two adjacent QDs, say QDi and QDj, described by a direct Coulomb matrix element $V_{ij}$ in a Hubbard model for a QD array is represented by the capacitive coupling of $C_{Mij}$ in the capacitance network model.[8,40,46] Here, the capacitance $C_{M14}$ between QD1 and QD4 is assumed to be small enough to be ignored, since the Coulomb interaction between the electrons in the two QDs is very weak due to the fact that the two QDs are separated by two other QDs. The capacitance $C_{G3\text{-}QD4}$ ($C_{G7\text{-}QD1}$) is also negligible owing to the long distance between gate G3 (G7) and QD4 (QD1). Based on the orthodox theory, we calculate the charge stability diagram of the QQD device (see details in Part IV of Supplementary Materials) and show the results of calculations in Fig. 4(b). The results depict a good agreement with our experimental data and thus confirm our assignments of the current lines observed in the measured charge stability diagram of Fig. 3(a).

*3.4. Tuning the InAs nanowire QQD to a regime where all the four QDs are close to resonance with the Fermi level of the source and drain electrodes*

Fig. 5 shows an experimental building-up of a regime of interest, where the energy levels of all the four QDs in the QQD are energetically close to resonance with the Fermi level of the source and drain electrodes, by adjusting the electrostatic potentials of QD2 and QD3 with gates G4 and G6. Here, we focus on an area of the charge stability diagram marked by the white rectangle in Fig. 3(a), where four adjacent current lines belonging to the four resonant QDs are present. Fig. 5(a) displays the charge stability diagram measured at $V_{G4} = -0.384$ V and $V_{G6} = -1.046$ V, where the grey, yellow and green dashed lines are associated with resonant transport through QD2, QD3 and QD4, respectively. We then gradually alter $V_{G6}$ from $-1.046$ V towards more negative values. As a result, the position of the current line of QD3 moves up gradually, while the position of the current line of QD4 remains almost unchanged. At $V_{G6} = -1.0535$ V, the spacing between these two current lines reaches a minimum as shown in Fig. 5(b) (details in Part V of Supplementary Materials). At the minimum spacing, QD3 and QD4 can be approximately regarded as on resonance with the Fermi level of the source and drain electrodes. At the same time, we notice that the current line marked by the grey dashed line arising from QD2 intersects with the current lines arising from QD3 and QD4 in the area marked by the white rectangle in Fig. 5(b). When we regard QD2, QD3 and QD4 as a TQD out of the QQD, this specific area corresponds to a resonant regime where all the three QDs are close to resonance with the Fermi level of the source and drain electrodes as illustrated in Fig. 5(e).[8] Fig. 5(c) displays the charge stability diagram of the QQD at $V_{G4} = -0.3843$ V and $V_{G6} = -1.0535$ V. Here, a current line marked with a blue dashed line, arising from QD1 on resonance, moves in and intersects with the current lines arising from QD3 and QD4, see the area marked by the red rectangle, where QD1, QD3 and QD4 are close to being on resonance with the Fermi level of the source and drain electrodes, as illustrated in Fig. 5(f). Now, when $V_{G4}$ is tuned gradually towards more negative values, the current line of QD2 moves to the right with the position of the current line of QD1 being approximately unchanged. As a consequence, when $V_{G4}$ is changed to

−0.386 V, the spacing between the current lines of QD2 and QD1 reaches a minimum, as shown in Fig. 5(d). The features of the current lines in the region marked by the red rectangle are a transport characteristic of the linear QQD in a regime where all the four QDs are close to being on resonance with the Fermi level of the source and drain electrodes, as illustrated in Fig. 5(g). These current line features in this nearly all-QD-resonant condition can be well reproduced by the simulation using the circuit model shown in Fig. 4(a) (see Part IV of Supplementary Materials). The successful realizations of the three regimes, where two QDs, three QDs and all the four QDs in the QQD are close to being on resonance with the Fermi level of source and drain electrodes, respectively, imply that the nanowire QQD possesses an excellent tunability and could work in more exotic regimes for quantum computation and simulation processes.

## 4. Conclusions

In summary, a linear QQD is realized in an InAs nanowire via fine finger gate technique and is studied by two-dimensional charge stability diagram measurements. Four characteristic groups of resonant current lines of different slopes are observed in the measured charge stability diagrams and are identified as arising from resonant tunneling through different QDs. It is also demonstrated that the energy levels of each QD can be sensitively controlled by a desired plunger gate, showing a superior gate tunability and thus allowing to efficiently bring the QQD to the regimes where two QDs, three QDs, and all the four QDs are energetically close to being on resonance with the Fermi level of the source and drain electrodes. A capacitance network model has also been developed and the simulated charge stability diagrams based on the model show a good agreement with the experiments. This work indicates that multiple QDs formed in InAs nanowires via finger gates could be a versatile platform to achieve integrated qubits and to perform quantum simulations.

## Acknowledgements

This work is supported by the Ministry of Science and Technology of China through the National Key Research and Development Program of China (Grant Nos. 2017YFA0303304, 2016YFA0300601, 2017YFA0204901, and 2016YFA0300802), the National Natural Science Foundation of China (Grant Nos. 91221202, 91421303, 11874071, 11974030 and 61974138), the Beijing Academy of Quantum Information

Sciences (No. Y18G22), and the Beijing Natural Science Foundation (Grant Nos. 1202010 and 1192017). DP also acknowledges the support from Youth Innovation Promotion Association, Chinese Academy of Sciences (No. 2017156).

## CAPTIONS

**Fig. 1** (a) SEM image (in false color) of the QQD device studied in this work. An array of finger gates made of 5-nm-thick Ti and 10-nm-thick Au with a gate width of ~30 nm and a pitch of ~70 nm are fabricated on a heavily p-doped Si substrate, which is capped with a 200-nm-thick layer of $SiO_2$. The finger gate array is covered by a 10-nm-thick layer of $HfO_2$. A nanowire of ~30 nm in diameter is then placed on top and is contacted by the source (S) and drain (D) electrodes made of 5-nm-thick Ti and 90-nm-thick Au. Two wide metal stripes which are made in the same time as the finger gate array are used to support the nanowire and the contact electrodes. (b) Cross-sectional schematic view of the device, where a linear QQD formed in the InAs nanowire using the finger gates underneath is indicated, and measurement circuit setup.

**Fig. 2** (a) Source-drain current $I_{ds}$ of the left DQD formed in the InAs nanowire on top of gates G1 to G5 measured at $V_{ds}$ = 0.1 mV as a function of gate voltages $V_{G3}$ and $V_{G4}$. The voltages applied to gates G1 and G5 are set at $V_{G1}$ = −2.56 V and $V_{G5}$ = −3.38 V. Gate G3 is used both to define the inter-dot tunneling barrier and to tune the charge states in the two QDs. Gate G2 is unused and is left floating. (b) Source-drain current $I_{ds}$ of the right DQD formed in the nanowire on top of gates G5 to G9 measured at $V_{ds}$ = 0.1 mV as a function of gate voltages $V_{G6}$ and $V_{G8}$. The voltages applied to barrier gates G5, G7, and G9 are set at $V_{G5}$ = −3.40 V, $V_{G7}$ = −3.65 V, and $V_{G9}$ = −3.86 V, respectively. (c) Schematic illustration of the left DQD. (d) Schematic illustration of the right DQD.

**Fig. 3** (a) Characteristic charge stability diagram of the InAs nanowire QQD as shown in Fig. 1(a), where the source-drain current $I_{ds}$ measured at $V_{ds}$ = 0.2 mV is plotted as a function of $V_{G3}$ and $V_{G7}$. The QQD is defined on top of gates G1 to G9 with the voltages applied to barrier gates G1, G5, and G9 set at $V_{G1}$ = −2.55 V, $V_{G5}$ = −3.35 V, and $V_{G9}$ = −3.855 V, and the voltages applied to plunger gates G4, G6 and G8 set at $V_{G4}$ = −0.38 V, $V_{G6}$= −1.05 V and $V_{G8}$ = −0.94 V. Gates G3 and G7 are used as tuning gates, while gate G2 is unused (i.e., left floating) throughout the measurements. Resonant current lines observed in the charge stability diagram are categorized into four groups, which are labeled as groups 1 to 4, according to their slopes and are marked by dashed lines of four different colors. As described in the text, the current lines marked by blue, grey, yellow, and green dashed lines represent the resonant current lines arising from tunneling via the energy levels of QD1, QD2, QD3, and QD4, respectively. Labels I to IV mark four representative cases at which two different groups of resonant current lines cross or show an avoiding crossing. (b) Energy level diagram corresponding to a current line marked by a blue dashed line in (a). (c) Energy level diagram corresponding to case I, where the energy levels of QD1 and QD3 are on resonance with the Fermi level of the source and drain electrodes. (d) The same as in (a) but only the right half of the charge stability diagram is shown with $V_{G6} = -1.05$ V clearly indicated. (e) and (f) The same as (d) but for $V_{G6} = -1.046$ V and $V_{G6} = -1.042$ V, respectively. Two current lines in group 3, marked as lines M and N, are seen clearly to move toward more negative values of $V_{G7}$ with increasing $V_{G6}$.

**Fig. 4** (a) Equivalent circuit of the capacitance network model employed to simulate the charge stability diagram of the linear nanowire QQD device. In the circuit, the four QDs are marked with colored ovals. The voltages applied to the source and drain electrodes, two plunger gates (G4 and G6), and two barrier gates (G3 and G7) are explicitly marked. The other gate voltages are fixed during the charge stability diagram measurements and are therefore not explicitly included in the equivalent circuit. Their effects are included implicitly in setting of the initial electrostatic potentials of the QDs and the parameters of the relevant inter-dot couplings as well as the couplings to the source and drain electrodes. (b) Characteristic charge stability diagram of the QQD device obtained by the simulation based on the capacitance network model.

**Fig. 5** Tuning of the nanowire QQD to a regime where all the four QDs are energetically on resonance with the Fermi level of the source and drain electrodes. (a)-(d) Charge stability diagrams (i.e., $I_{ds}$ as a function of $V_{G3}$ and $V_{G7}$) measured for the QQD at $V_{G4} = -0.384$ V and $V_{G6} = -1.046$ V, $V_{G4} = -0.384$ V and $V_{G6} = -1.0535$ V, $V_{G4} = -0.3843$ V and $V_{G6} = -1.0535$ V, and $V_{G4} = -0.386$ V and $V_{G6} = -1.0535$ V, respectively. The measurements are performed by setting the barrier gate voltages at $V_{G1} = -2.55$ V, $V_{G5} = -3.35$ V and $V_{G9} = -3.855$ V, and the plunger gate voltage at $V_{G8} = -0.94$ V. The source-drain voltage $V_{ds}$ applied is 0.2 mV in (a) and 40 μV in the other three panels. The current lines marked by the blue, grey, yellow and green

dashed lines arise from resonant transport via the energy levels in QD1, QD2, QD3 and QD4, respectively. These figures show how the charge stability diagram has evolved in our searching for the all-QD-on resonance regime of the QQD by adjusting $V_{G4}$ and $V_{G6}$. (e) Energy level diagram of the QQD in the region marked by the white rectangle in (b). (f) Energy level diagram of the QQD in the region marked by the red rectangle in (c). (g) Energy level diagram of the QQD in the region marked by the red rectangle in (d).

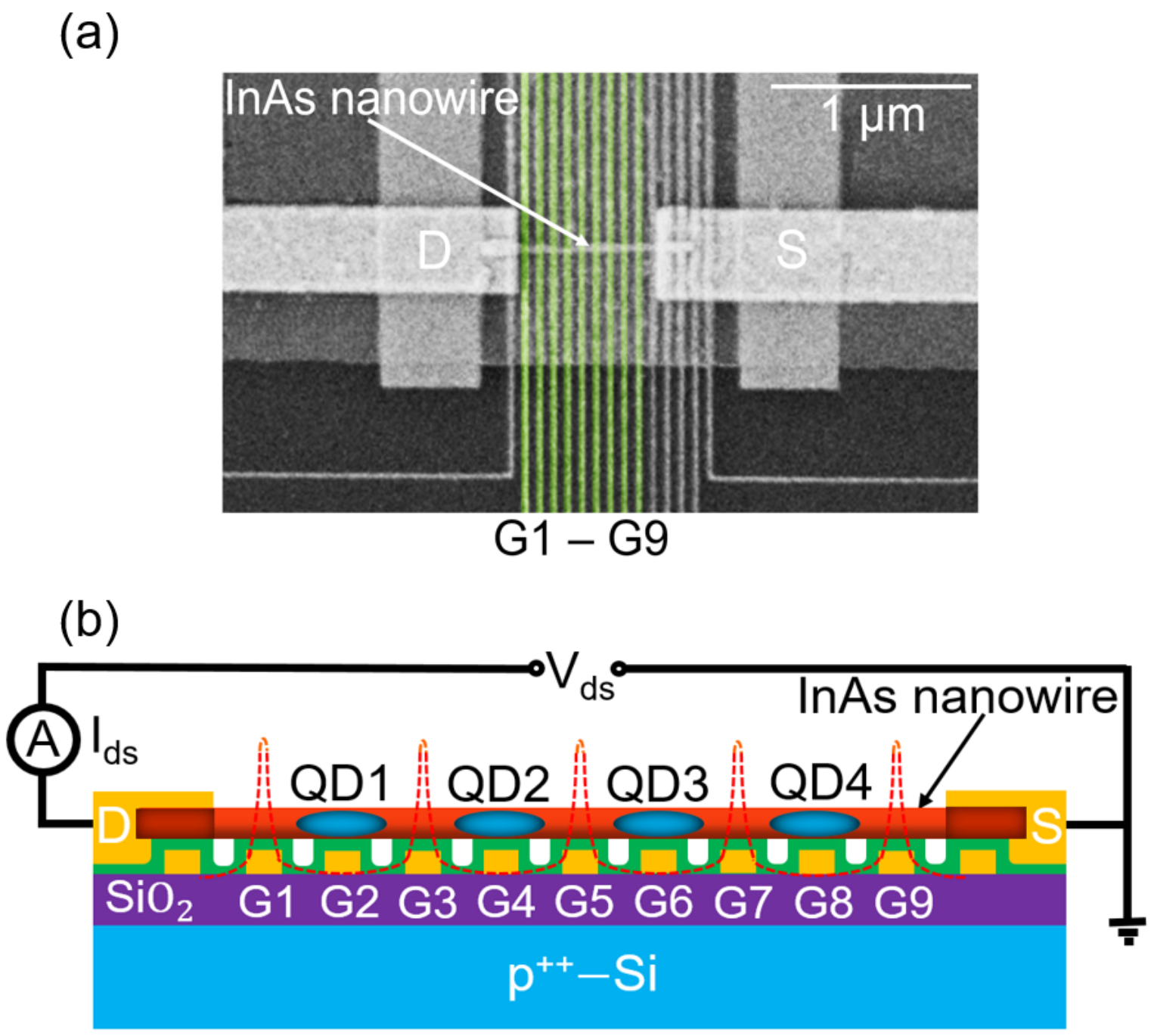

**Figure 1, by Jingwei Mu *et al*.**

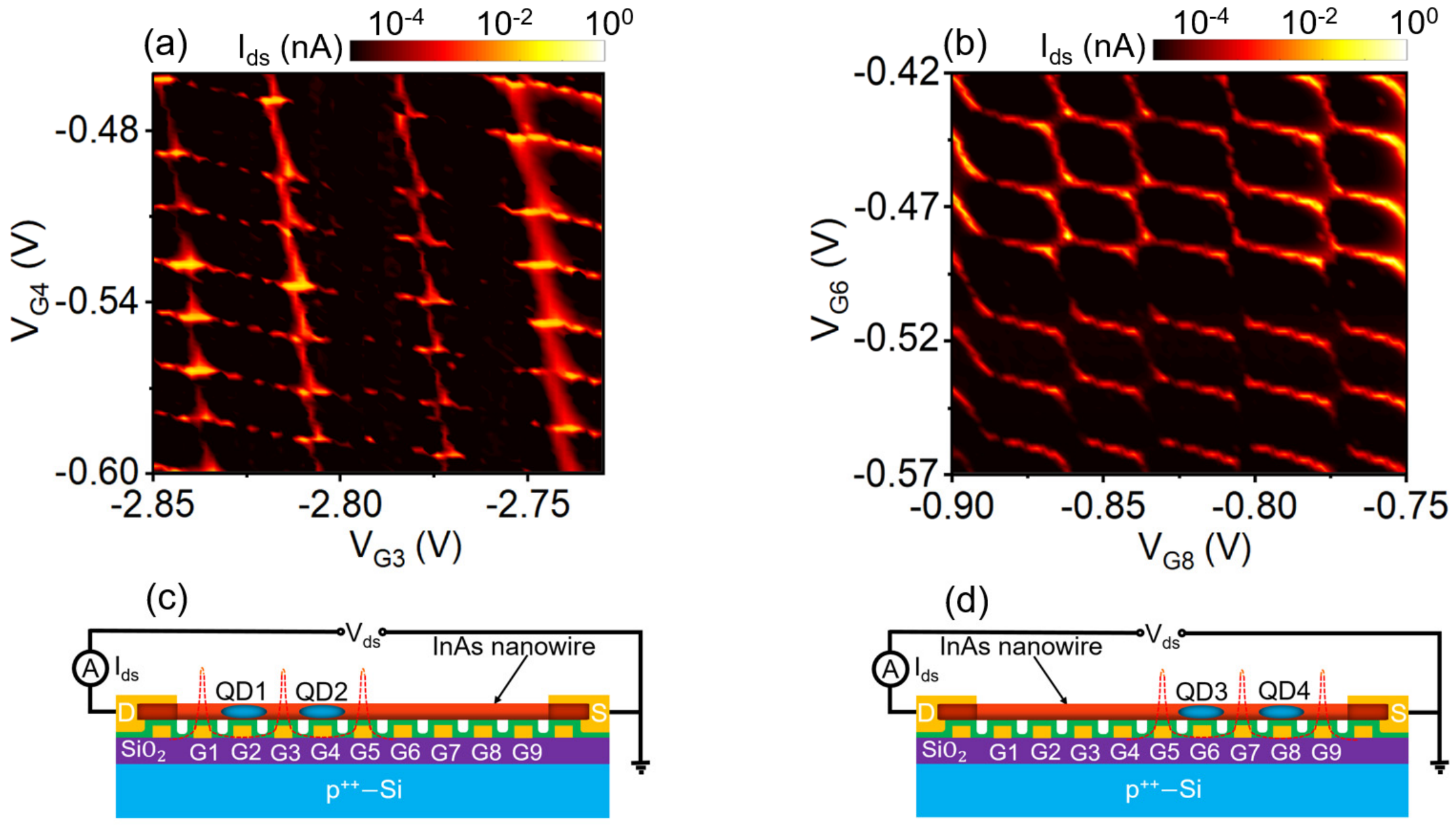

**Figure 2, by Jingwei Mu *et al*.**

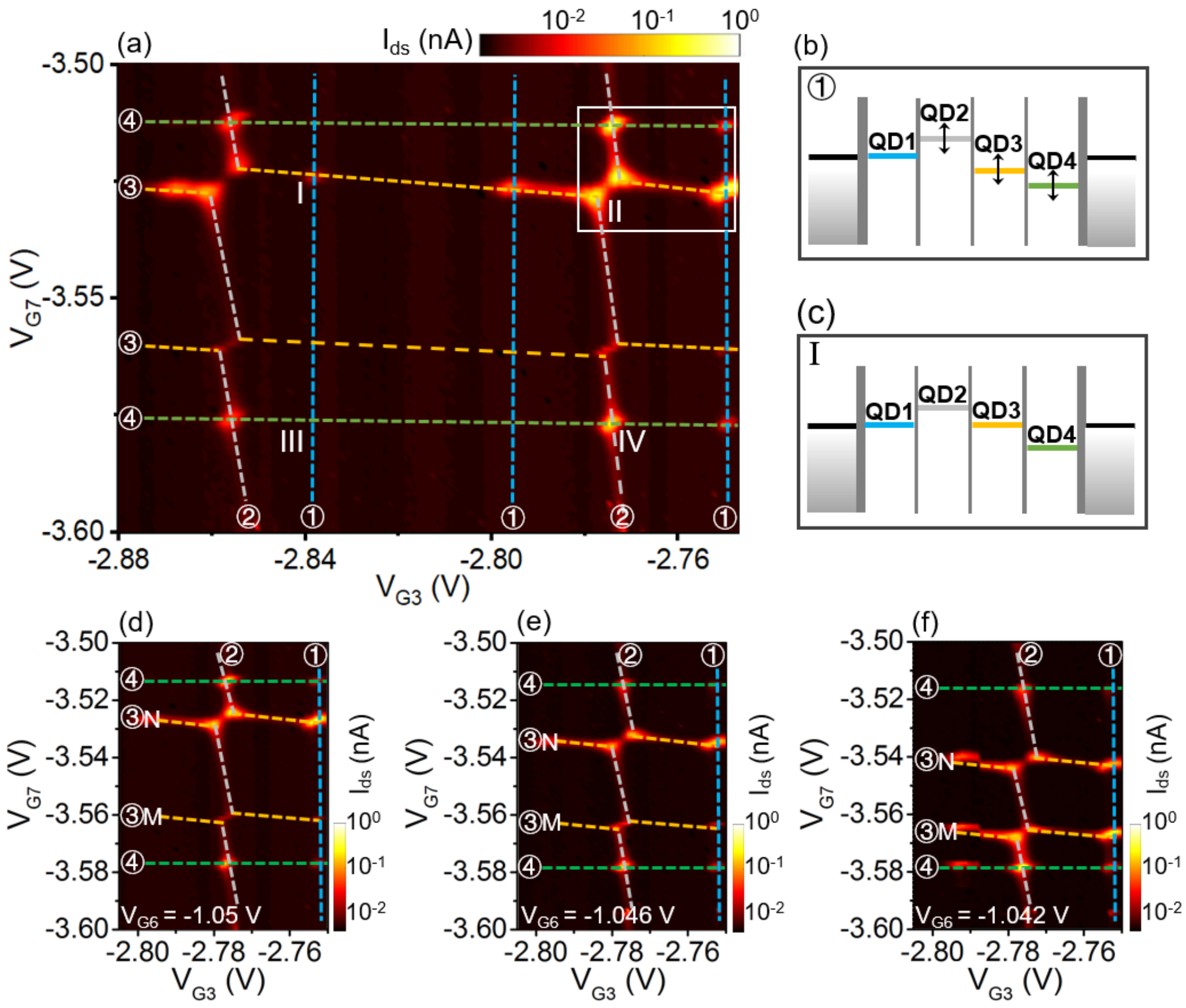

**Figure 3, by Jingwei Mu *et al*.**

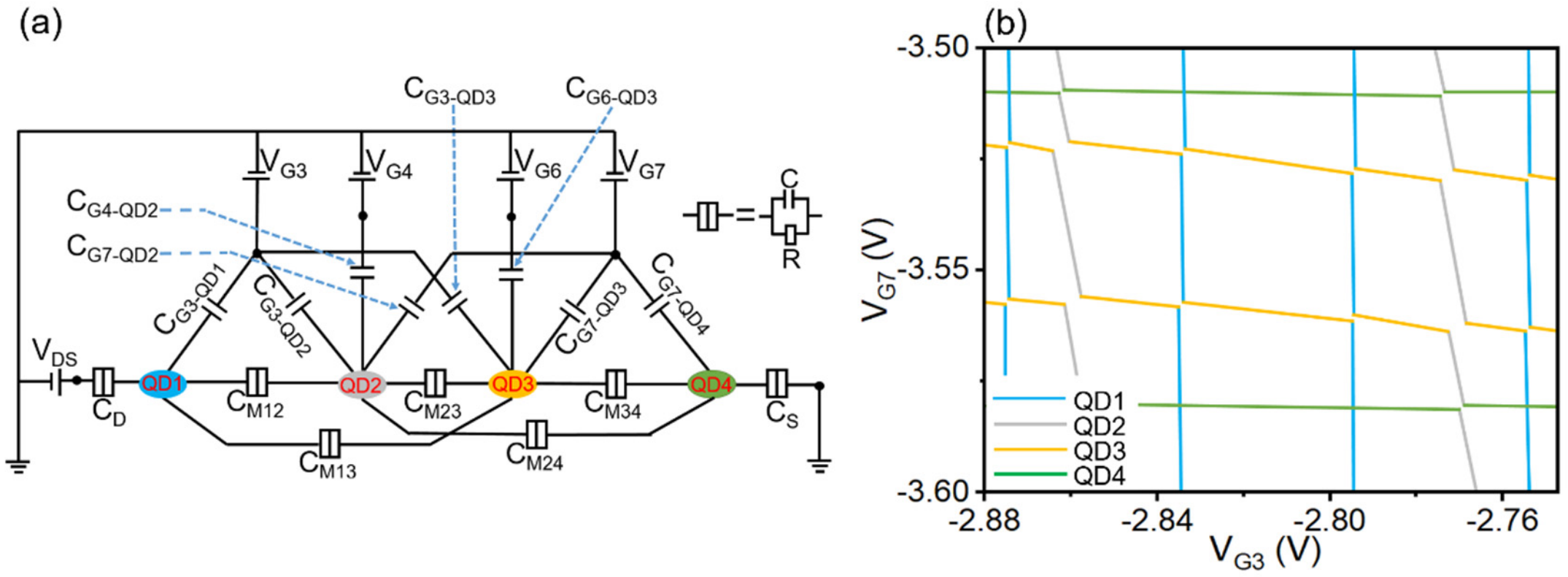

**Figure 4, by Jingwei Mu *et al*.**

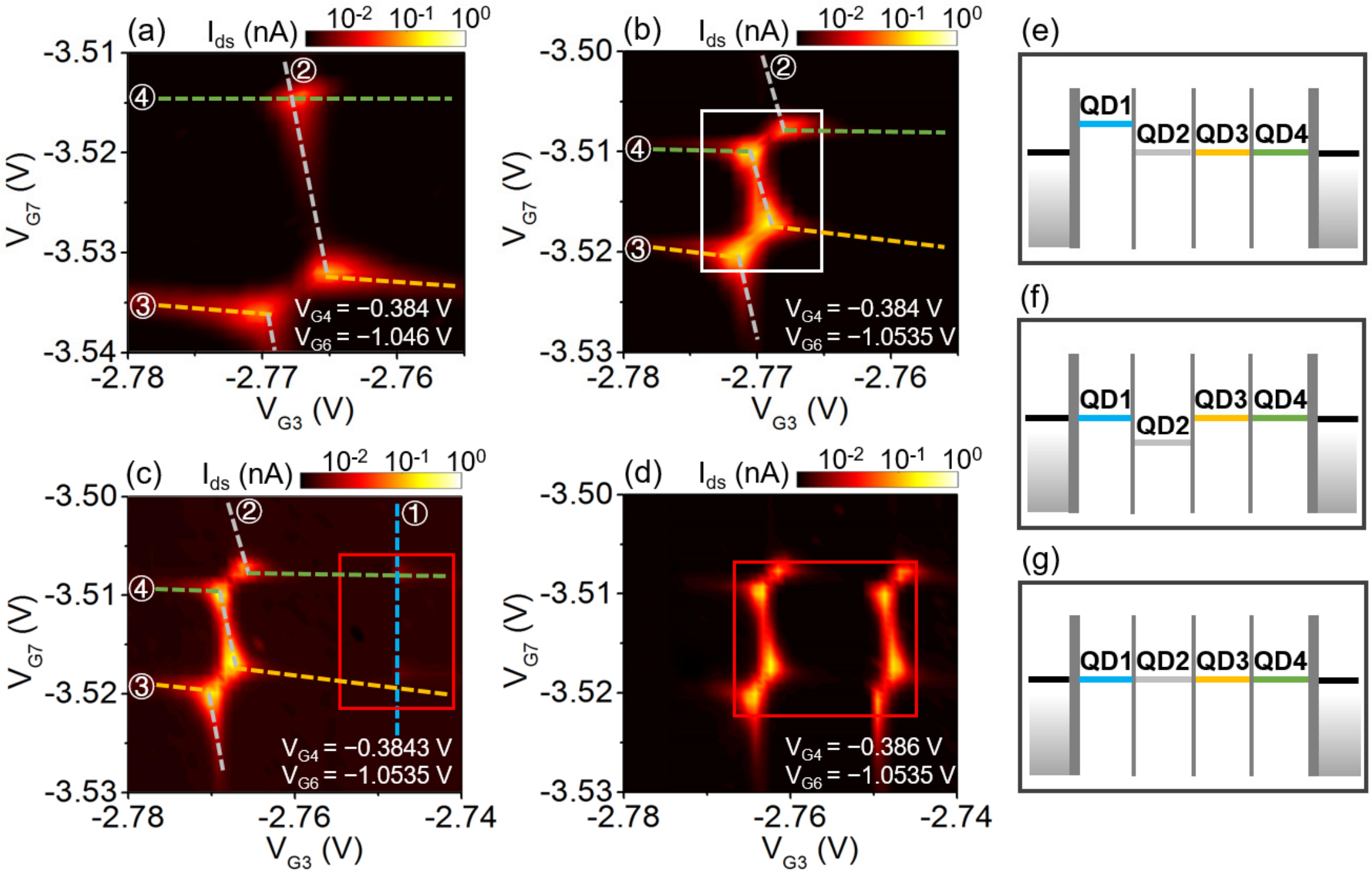

**Figure 5, by Jingwei Mu *et al*.**

# Supplementary Materials for

## Highly tunable quadruple quantum dot in a narrow bandgap semiconductor InAs nanowire

Jingwei Mu,[a,c] Shaoyun Huang,[a] Zhi-Hai Liu,[a] Weijie Li,[a,c] Ji-Yin Wang,[a] Dong Pan,[b] Guang-Yao Huang,[a] Yuanjie Chen,[a] Jianhua Zhao,[b,d,†] and H. Q. Xu[a,c,d,*]

[a]*Beijing Key Laboratory of Quantum Devices, Key Laboratory for the Physics and Chemistry of Nanodevices and Department of Electronics, Peking University, Beijing 100871, China*

[b]*State Key Laboratory of Superlattices and Microstructures, Institute of Semiconductors, Chinese Academy of Sciences, P.O. Box 912, Beijing 100083, China*

[c]*Academy for Advanced Interdisciplinary Studies, Peking University, Beijing 100871, China*

[d]*Beijing Academy of Quantum Information Sciences, Beijing 100193, China*

Corresponding authors. Emails: *hqxu@pku.edu.cn; †jhzhao@semi.ac.cn

(January 18, 2020)

### Outline

**I. Charge stability diagram of a single QD defined in the InAs nanowire**

**II. Complementary measurements of the QQD device studied in the main article**

**III. Verification of the current lines that arise from resonant transport through QD2 and QD4 in the QQD device studied in the main article**

**IV. Electrostatic capacitance network model for the QQD device studied in the main article**

**V. Achieving the resonance between the energy levels of QD3 and QD4 in the QQD device studied in the main article**

## I. Charge stability diagram of a single QD defined in the InAs nanowire

Figure S1 displays the differential conductance $dI_{ds}/dV_{ds}$ of a single quantum dot (QD) defined in the InAs nanowire as a function of source-drain bias voltage $V_{ds}$ and voltage $V_{G4}$ applied to plunger gate G4 (charge stability diagram). Here, gates G3 and G5 are used to define the two tunneling barriers of the single QD. The regular Coulomb diamonds as well as the close points seen at a zero $V_{ds}$ between neighboring Coulomb diamonds indicate the formation of a single QD. The small electron effective mass of InAs, $m_e^*$=0.23$m_e$, leads to strong quantum confinement effect[1] and QDs built from InAs nanowires can easily possess a large energy level separation. Thus, a quantization energy of ~1.5 meV is observed in the measured charge stability diagram of the single QD. Besides, we also extract a single electron charging energy of $E_C$~5 meV in the QD from the measured charge stability diagram.

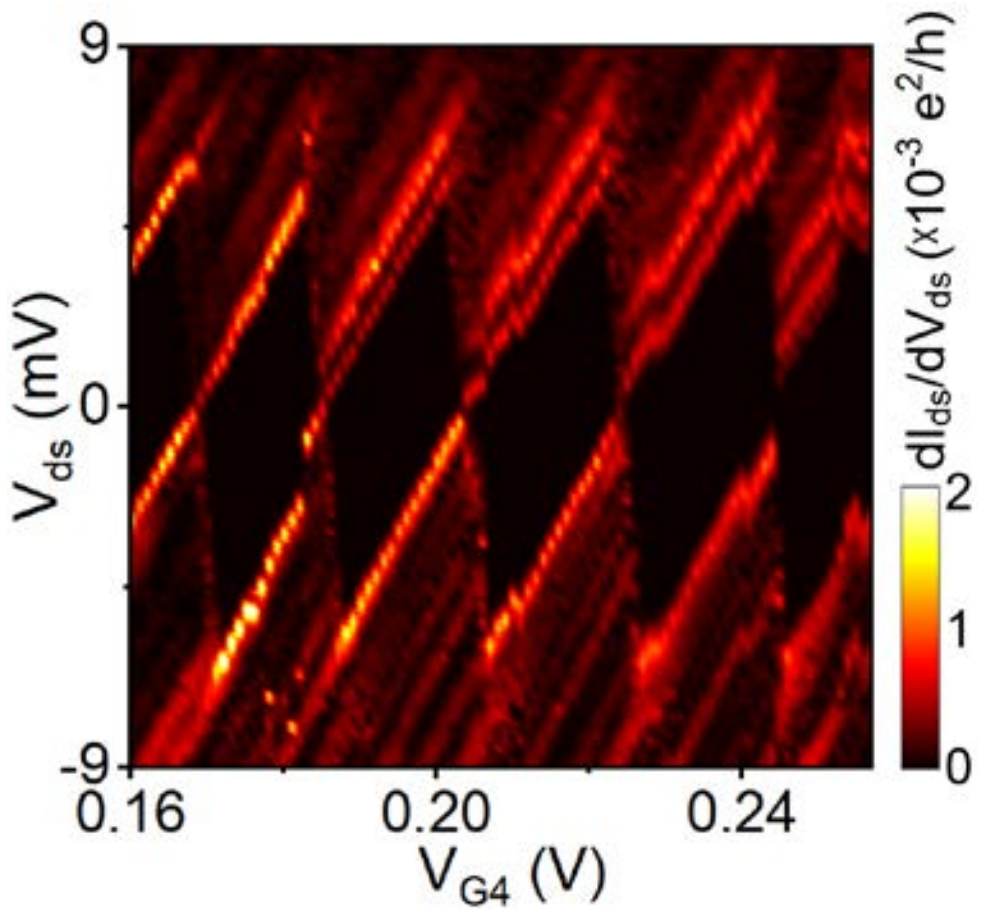

**Fig. S1** Differential conductance $dI_{ds}/dV_{ds}$ of a single QD defined in the InAs nanowire as a function of source-drain bias voltage $V_{ds}$ and gate voltage $V_{G4}$. Here, the single QD is defined by setting the voltages applied to gates G3 and G5 at $V_{G3} = -2.85$ V and $V_{G5} = -3.5$ V, see Fig. 1 in the main article for the device structure and the measurement circuit setup.

## II. Complementary measurements of the QQD device studied in the main article

Figure S2 shows the source-drain current $I_{ds}$ measured for the quadruple quantum dot (QQD) defined in the InAs nanowire (see Fig. 1 in the main article) at $V_{ds}$ = 0.1 mV as a function of voltages $V_{G4}$ and $V_{G6}$ applied to plunger gates G4 and G6 [Fig. S2(a)], and as a function of voltages $V_{G4}$ and $V_{G8}$ applied to plunger gates G4 and G8 [Fig. S2(b)]. Both charge stability diagrams show the similar characteristics of a double QD (DQD), i.e., only two groups of resonant current lines of different slopes are present in each of the two charge stability diagrams.[2] Thus, we can infer that the current lines in Fig. S2(a) [Fig. S2(b)] arise from resonant electron transport through QD2 and QD3 (QD2 and QD4), and plunger gates G4, G6 and G8 couple strongly to their locally addressed QDs but weakly to the other QDs. In other words, the parasitical capacitances between a plunger gate and QDs other than its locally addressed one are very small in the QQD device. Therefore, it is difficult to map out the complete charge state configuration of the QQD in a two-dimensional charge stability diagram effectively by scanning plunger gates. This is the reason why in the main article a two-dimensional charge stability diagram of the QQD is measured and plotted as a function of gate voltages $V_{G3}$ and $V_{G7}$. It is important to note that, in Fig. S2(b), no recognizable avoiding crossing of two current lines of different slopes is found, due to the weak direct coupling between QD2 and QD4, which are separated by QD3.

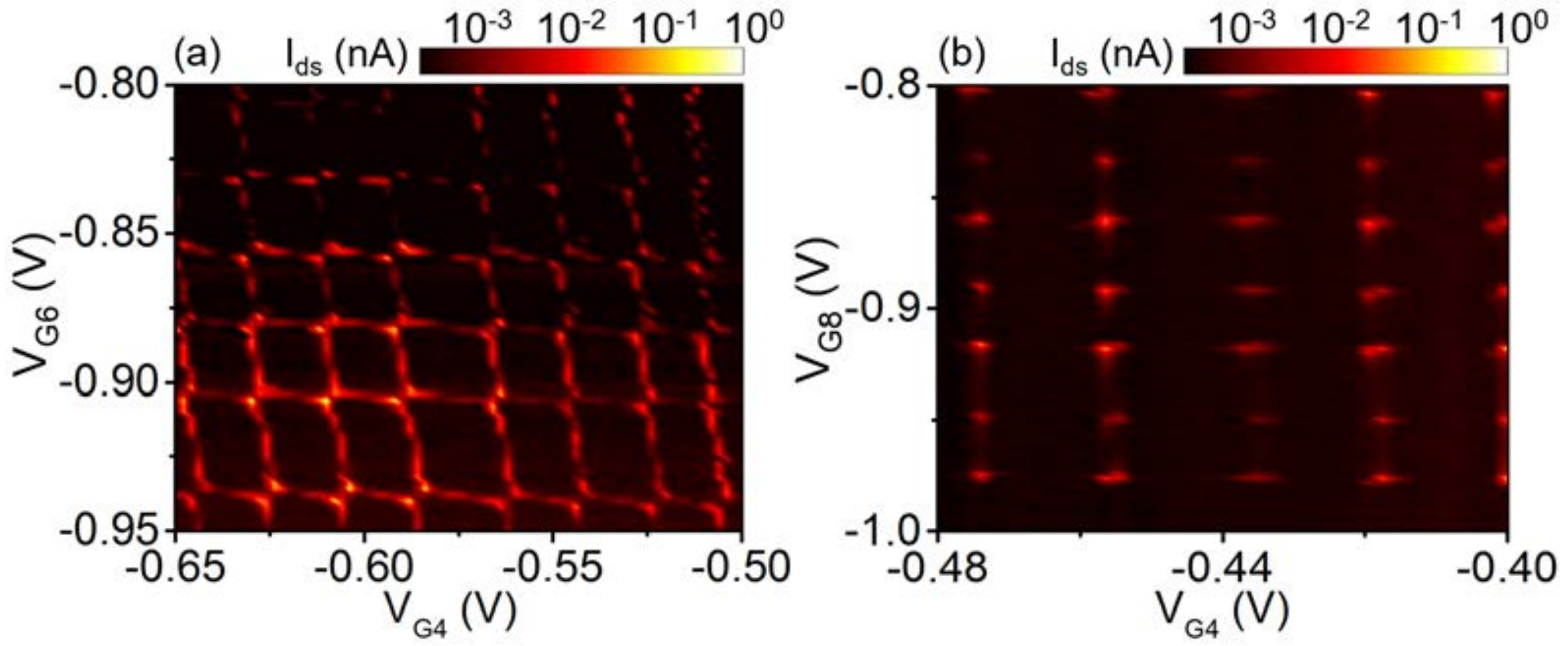

**Fig. S2** (a) Source-drain current $I_{ds}$ measured at $V_{ds}$ = 0.1 mV for the QQD device defined in the InAs nanowire as shown in Fig. 1 of the main article as a function of voltages $V_{G4}$ and $V_{G6}$ applied to plunger gates G4 and G6, while setting the voltage applied to plunger gate G8 at $V_{G8}$ = −1 V. (b) the same as (a) but measured as a function of voltages $V_{G4}$ and $V_{G8}$ applied to plunger gates G4 and G8, while setting the voltage applied to plunger gate G6 at $V_{G6}$ = −0.6 V. The QQD is defined in the InAs nanowire with the voltages applied to barrier gates G1, G3, G5, G7, and G9 set at $V_{G1}$ = −2.545 V, $V_{G3}$ = −2.74 V, $V_{G5}$ = −3.39 V, $V_{G7}$ = −3.65 V, and $V_{G9}$ = −3.845 V, respectively. Plunger gate G2 is unused and is left floating throughout the measurements.

## III. Verification of the current lines that arise from resonant transport through QD2 and QD4 in the QQD device studied in the main article

Plunger gate G8 possesses a strong coupling to its locally addressed QD4 but relatively weak cross-talks to the other QDs. We, therefore, tune the voltage $V_{G8}$ applied to plunger gate G8 and examine the changes of current lines in the measured charge stability diagrams shown in Fig. S3(a) and S3(b). Here, we observe that the positions of the current lines in groups 1, 2 and 3 remain approximately unchanged, while the positions of the current lines in group 4 move toward the positive direction of $V_{G7}$, when the voltage $V_{G8}$ is tuned from $V_{G8}$ = −0.94 V to $V_{G8}$ = −0.95 V.

Thus, the current lines in group 4 evidently arise from resonant electron transport through QD4. It should be noted that a current line in group 3, marked by P in Fig. S3(a), moves noticeably also towards the positive direction of $V_{G7}$. This is because an energy level of QD4 gradually approaches the energy level of QD3 represented by current line P and pushes the energy level to move due to level repulsion with decreasing $V_{G8}$. In the same way, the assignment of the current lines in group 2 as arising from resonant transport through QD2 is double-checked by tuning the voltage $V_{G4}$ applied to plunger gate G4 from $V_{G4} = -0.380$ V to $V_{G4} = -0.385$ V and examining the changes of current lines in the measured charge stability diagrams as shown in Fig. S3(c) and S3(d). Clearly, it is seen that the current lines in group 2 move significantly towards the positive direction of $V_{G3}$ and all the other current lines remain approximately unchanged with decreasing $V_{G4}$, confirming our previous assignment for the current lines in group 2.

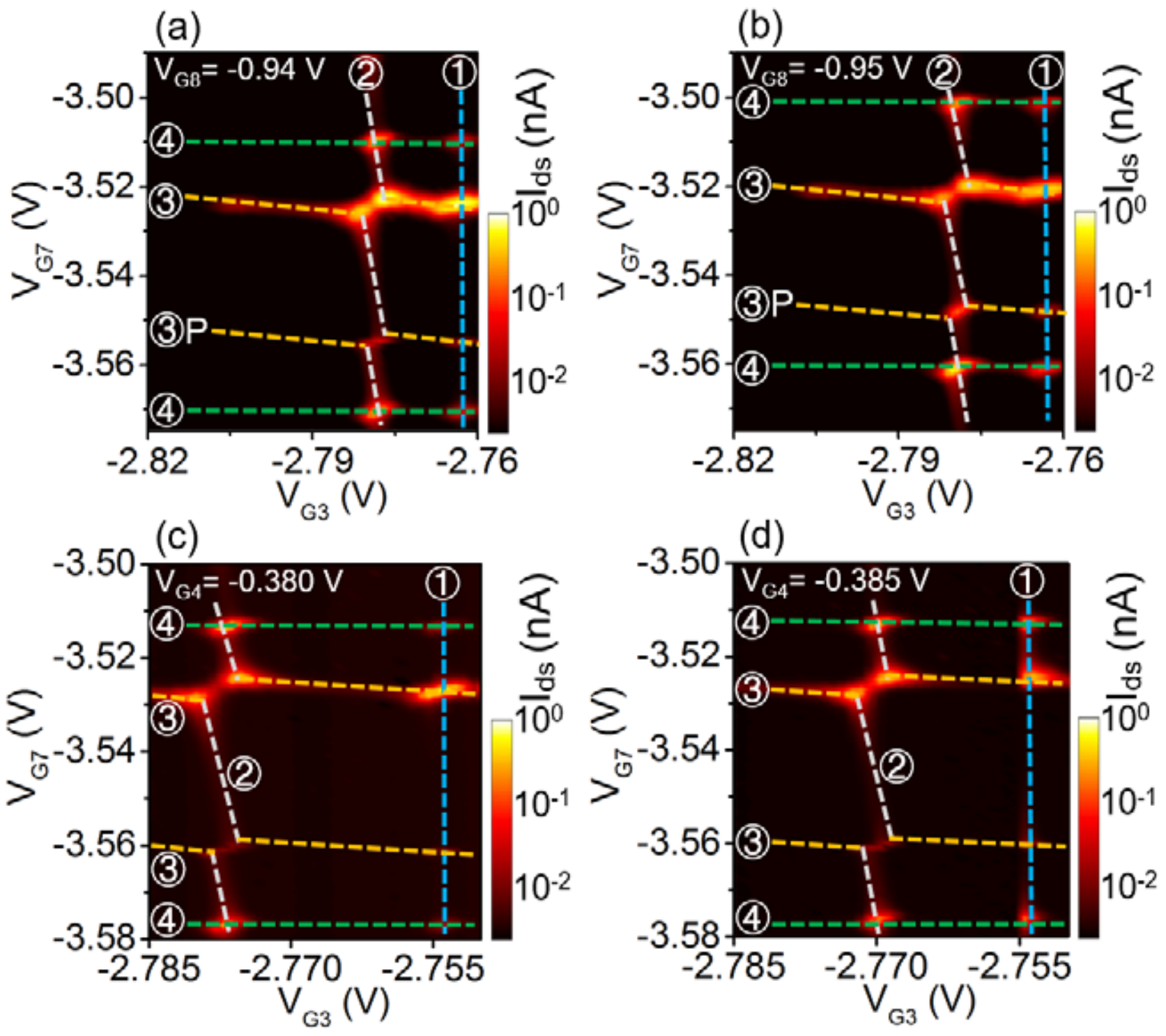

**Fig. S3** (a) and (b) Source-drain current $I_{ds}$ measured for the InAs nanowire QQD as shown in Fig. 1 of the main article at $V_{ds}$ = 0.2 mV as a function of $V_{G3}$ and $V_{G7}$ at two different values of plunger gate voltage $V_{G8}$. Here, to define the InAs nanowire QQD, the voltages applied to barrier gates G1, G5, and G9 are set at $V_{G1} = -2.538$ V, $V_{G5} = -3.35$ V, and $V_{G9} = -3.855$ V, and the voltages applied to plunger gates G4 and G6 are set at $V_{G4} = -0.38$ V and $V_{G6} = -1.05$ V. (c) and (d) The same as (a) and (b) but for the measurements at $V_{ds}$ = 0.1 mV and at two different values of $V_{G4}$. Here, the QQD is defined by setting the voltages applied to barrier gates G1, G5, and G9 at $V_{G1} = -2.55$ V, $V_{G5} = -3.35$ V, and $V_{G9} = -3.855$ V, and the voltages applied to plunger gates G6 and G8 at $V_{G6} = -1.05$ V and $V_{G8} = -0.94$ V.

## IV. Electrostatic capacitance network model for the QQD device studied in the main article

Electrostatic capacitance network models have been successfully employed to describe the characteristics of the charge stability diagrams of a DQD and a Triple QD (TQD).[2-5] Here, we present an electrostatic capacitance network model for the InAs nanowire QQD studied in the main article. In our work, the charge stability diagrams of the QQD are measured at different voltages $V_{G4}$ and $V_{G6}$ as a function of $V_{G3}$ and $V_{G7}$ (cf. Fig. 1 in the main article). The QQD device can modeled by a circuit shown in Fig. 4(a) of the main article, which includes the source and drain electrodes, four gates (G3, G4, G6 and G7) that are changed in the measurements, and four QDs (QD1, QD2, QD3 and QD4) that act as four nodes and marked with colored ovals in the model circuit. Based on the classical theory, the charges on individual QDs can be expressed in terms of relevant gate voltages and capacitances as,

$$
\begin{aligned}
Q_{QD1} &= C_S(V_{QD1} - V_S) + C_{G3\text{-}QD1}(V_{QD1} - V_{G3}) + C_{M12}(V_{QD1} - V_{QD2}) + C_{M13}(V_{QD1} - V_{QD3}), \\
Q_{QD2} &= C_{M12}(V_{QD2} - V_{QD1}) + C_{G3\text{-}QD2}(V_{QD2} - V_{G3}) + C_{G4\text{-}QD2}(V_{QD2} - V_{G4}) + C_{G7\text{-}QD2}(V_{QD2} - V_{G7}) \\
&\quad + C_{M23}(V_{QD2} - V_{QD3}) + C_{M24}(V_{QD2} - V_{QD4}), \\
Q_{QD3} &= C_{M23}(V_{QD3} - V_{QD2}) + C_{G6\text{-}QD3}(V_{QD3} - V_{G6}) + C_{G7\text{-}QD3}(V_{QD3} - V_{G7}) + C_{G3\text{-}QD3}(V_{QD3} - V_{G3}) \\
&\quad + C_{M13}(V_{QD3} - V_{QD1}) + C_{M34}(V_{QD3} - V_{QD4}), \\
Q_{QD4} &= C_{M34}(V_{QD4} - V_{QD3}) + C_{G7\text{-}QD4}(V_{QD4} - V_{G7}) + C_{M24}(V_{QD4} - V_{QD2}) + C_D(V_{QD4} - V_D). \qquad (1)
\end{aligned}
$$

Defining the vectors

$$V_{QD} = \{V_{QD1}, V_{QD2}, V_{QD3}, V_{QD4}\}^T, \quad V_G = \{V_S, V_{G3}, V_{G4}, V_{G6}, V_{G7}, V_D\}^T, \quad Q_{QD} = \{Q_1, Q_2, Q_3, Q_4\}^T,$$

Eq. (1) can be written as

$$Q_{QD} = C_{QD} \cdot V_{QD} + C_{QD\text{-}G} \cdot V_G, \qquad (2)$$

with matrices $C_{QD}$ and $C_{QD\text{-}G}$ given by

$$C_{QD} = \begin{pmatrix} C_1 & -C_{M12} & -C_{M13} & 0 \\ -C_{M12} & C_2 & -C_{M23} & -C_{M24} \\ -C_{M13} & -C_{M23} & C_3 & -C_{M34} \\ 0 & -C_{M24} & -C_{M34} & C_4 \end{pmatrix}, \qquad (3)$$

$$C_{QD\text{-}G} = \begin{pmatrix} -C_S & -C_{G3\text{-}QD1} & 0 & 0 & 0 & 0 \\ 0 & -C_{G3\text{-}QD2} & -C_{G4\text{-}QD2} & 0 & -C_{G7\text{-}QD2} & 0 \\ 0 & -C_{G3\text{-}QD3} & 0 & -C_{G6\text{-}QD3} & -C_{G7\text{-}QD3} & 0 \\ 0 & 0 & 0 & 0 & -C_{G7\text{-}QD4} & -C_D \end{pmatrix}, \qquad (4)$$

where

$$C_1 = C_S + C_{G3\text{-}QD1} + C_{M12} + C_{M13}\,,$$

$$C_2 = C_{M12} + C_{G3\text{-}QD2} + C_{G4\text{-}QD2} + C_{M23} + C_{M24} + C_{G7\text{-}QD2}\,,$$

$$C_3 = C_{M23} + C_{G6\text{-}QD3} + C_{G7\text{-}QD3} + C_{M13} + C_{M34} + C_{G3\text{-}QD3}\,,$$

$$C_4 = C_{M34} + C_{G7\text{-}QD4} + C_{M24} + C_D\,. \tag{5}$$

Then, the electric potentials on the QDs can be calculated from

$$V_{QD} = C_{QD}^{-1}\left(Q_{QD} - C_{QD\text{-}G}\cdot V_G\right), \tag{6}$$

and the electrostatic energy of the QQD reads

$$\begin{aligned} E &= \frac{1}{2}V_{QD}^T \cdot C_{QD} \cdot V_{QD} \\ &= \frac{1}{2}V_{QD}^T\left(Q_{QD} - C_{QD\text{-}G}\cdot V_G\right) \\ &= \frac{1}{2}\left(Q_{QD} - C_{QD\text{-}G}\cdot V_G\right)^T [C_{QD}^{-1}]^T\left(Q_{QD} - C_{QD\text{-}G}\cdot V_G\right). \end{aligned} \tag{7}$$

Because $Q_i = -eN_i$ ($i$ = 1, 2, 3, 4), the chemical potentials $\mu_i$ in individual QDs are given by

$$\mu_1(N_1,N_2,N_3,N_4,V_{G3},V_{G4},V_{G6},V_{G7}) = E(N_1,N_2,N_3,N_4,V_{G3},V_{G4},V_{G6},V_{G7}) - E(N_1-1,N_2,N_3,N_4,$$
$$V_{G3},V_{G4},V_{G6},V_{G7})\,,$$

$$\mu_2(N_1,N_2,N_3,N_4,V_{G3},V_{G4},V_{G6},V_{G7}) = E(N_1,N_2,N_3,N_4,V_{G3},V_{G4},V_{G6},V_{G7}) - E(N_1,N_2-1,N_3,N_4,$$
$$V_{G3},V_{G4},V_{G6},V_{G7})\,,$$

$$\mu_3(N_1,N_2,N_3,N_4,V_{G3},V_{G4},V_{G6},V_{G7}) = E(N_1,N_2,N_3,N_4,V_{G3},V_{G4},V_{G6},V_{G7}) - E(N_1,N_2,N_3-1,N_4,$$
$$V_{G3},V_{G4},V_{G6},V_{G7})\,,$$

$$\mu_4(N_1,N_2,N_3,N_4,V_{G3},V_{G4},V_{G6},V_{G7}) = E(N_1,N_2,N_3,N_4,V_{G3},V_{G4},V_{G6},V_{G7}) - E(N_1,N_2,N_3,N_4-1,$$
$$V_{G3},V_{G4},V_{G6},V_{G7})\,. \tag{8}$$

The electron occupancies of the four QDs can be manipulated by adjusting the voltages of the plunger gates or barrier gates. For single electron tunneling through QD*i* to occur, the chemical potentials $\mu_i$ in QD*i* at occupancies $N_i$ and $N_i + 1$ need to be equal. This can be achieved by adjusting a set of selected gate voltages while keeping the other gate voltages unchanged. In our experiment, we have measured single electron tunneling as a function of $V_{G3}$ and $V_{G7}$ at fixed values for the other gate voltages. Therefore, the condition for the single electron tunneling through QD*i* to take place is given by

$$\mu_i(N_i,V_{G3},V_{G7}) = \mu_i(N_i+1,V_{G3},V_{G7})\,. \tag{9}$$

In our experiment, we see that the current lines in each group, i.e., arising from resonant tunneling transport through a QD, are nearly parallel in the measured charge stability diagram of the QQD. Thus, the voltages differences $\Delta V_{G3\text{-}QDi}$ and $\Delta V_{G7\text{-}QDi}$, where $\Delta V_{G3\text{-}QDi}$ ($\Delta V_{G7\text{-}QDi}$) measures the distance in $V_{G3}$ ($V_{G7}$) between two neighboring current lines in group $i$ at a fixed value of $V_{G7}$ ($V_{G3}$), can be written as,

$$\mu_i(N_i, V_{G3}, V_{G7}) = \mu_i(N_i + 1, V_{G3} + \Delta V_{G3\text{-}QDi}, V_{G7})$$
$$\mu_i(N_i, V_{G3}, V_{G7}) = \mu_i(N_i + 1, V_{G3}, V_{G7} + \Delta V_{G7\text{-}QDi}) . \quad (10)$$

Note that, in the above equations, Eqs (9) and (10), we have omitted those occupancy numbers and gate voltages that have not been changed during the considered tunneling process. By solving Eq. (10) with the help of Eqs. (7) and (8), $\Delta V_{G3\text{-}QDi}$ and $\Delta V_{G7\text{-}QDi}$ can be obtained as,

$$\Delta V_{G3\text{-}QD1} = \frac{e}{C_{G3\text{-}QD1} + \frac{C_{M12}}{C_2} C_{G3\text{-}QD2} + (\frac{C_{M13}}{C_2} + \frac{C_{M12}C_{M23}}{C_2 C_3}) C_{G3\text{-}QD3}},$$

$$\Delta V_{G7\text{-}QD1} = \frac{e}{\frac{C_{M13}}{C_3} C_{G7\text{-}QD3} + (\frac{C_{M13}}{C_2} + \frac{C_{M12}C_{M23}}{C_2 C_3}) C_{G7\text{-}QD3}},$$

$$\Delta V_{G3\text{-}QD2} = \frac{e}{C_{G3\text{-}QD2} + \frac{C_{M12}}{C_1} C_{G3\text{-}QD1} + \frac{C_{M23}}{C_3} C_{G3\text{-}QD3}},$$

$$\Delta V_{G7\text{-}QD2} = \frac{e}{C_{G7\text{-}QD2} + \frac{C_{M23}}{C_3} C_{G7\text{-}QD3} + (\frac{C_{M24}}{C_4} + \frac{C_{M23}C_{M34}}{C_3 C_4}) C_{G7\text{-}QD4}},$$

$$\Delta V_{G3\text{-}QD3} = \frac{e}{C_{G3\text{-}QD3} + (\frac{C_{M13}}{C_1} + \frac{C_{M12}C_{M23}}{C_1 C_2}) C_{G3\text{-}QD1} + \frac{C_{M23}}{C_2} C_{G3\text{-}QD2}},$$

$$\Delta V_{G7\text{-}QD3} = \frac{e}{C_{G7\text{-}QD3} + \frac{C_{M34}}{C_4} C_{G7\text{-}QD4} + \frac{C_{M23}}{C_2} C_{G7\text{-}QD2}},$$

$$\Delta V_{G3\text{-}QD4} = \frac{e}{(\frac{C_{M24}}{C_2} + \frac{C_{M23}C_{M34}}{C_2 C_3}) C_{G3\text{-}QD2} + \frac{C_{M34}}{C_3} C_{G3\text{-}QD3}},$$

$$\Delta V_{G7\text{-}QD4} = \frac{e}{C_{G7\text{-}QD4} + \frac{C_{M34}}{C_3} C_{G7\text{-}QD3} + (\frac{C_{M24}}{C_2} + \frac{C_{M23}C_{M34}}{C_2 C_3}) C_{G7\text{-}QD2}} . \quad (11)$$

We take capacitances $C_{G3\text{-}QD1} \approx 4.10$ aF, $C_{G4\text{-}QD2} \approx 8.70$ aF and $C_{G6\text{-}QD3} \approx 6.91$ aF as extracted from the measurements of the two DQDs in Fig. 2 of the main article. Capacitances $C_S$ and $C_D$ are assumed to be equal and are estimated from the charge stability diagram of the single QD in Fig. S1 to be $C_S = C_D \approx 8.00$ aF. Because of the weak coupling between QD2 and QD4 (QD1 and QD3), the associated inter-dot capacitance is small and thus a value of $C_{M24} \approx 0.02$ aF ($C_{M13} \approx 0.02$ aF) is assumed. To extract the remaining capacitances in the QQD device, we consider the measured stability diagram shown in Fig. 3(a) of the main article. From this charge stability diagram, it is found that the individual voltage changes that are required to move from one resonant current line to the next line as a result of resonance transport through individual QDs are $\Delta V_{G3\text{-}QD1} = 0.042$ V, $\Delta V_{G3\text{-}QD2} = 0.083$ V, $\Delta V_{G7\text{-}QD2} = 0.448$ V, $\Delta V_{G3\text{-}QD3} = 0.461$ V, $\Delta V_{G7\text{-}QD3} = 0.034$ V, $\Delta V_{G7\text{-}QD4} = 0.063$ V, $\Delta V_{G7\text{-}QD1} \approx \infty$, and $\Delta V_{G3\text{-}QD4} \approx \infty$. Inserting these voltage change values into Eq. (11) and considering the fact that $C_{i=1,2,3,4} >> C_{Mij}$ (where, on the right side, $i<j$ with $i=1, 2, 3$ and $j=2, 3, 4$), all the remaining capacitance values can be extracted. All the capacitance parameters used to obtain the charge stability diagram of the QQD are listed in Table SI. Figure 4(b) of the main article displays a simulated charge stability diagram of the QQD based on the capacitance network model developed here with the input capacitance values listed in Table SI. It shows a good agreement with our experiment and thus confirm our assignments of the current lines observed in the measured charge stability diagram shown in Fig. 3(a) of the main article.

**TABLE SI.** Values of the capacitances, in units of aF, used to simulate the charge stability diagram of the nanowire QQD shown in Fig. 3(a) of the main article. The simulated charge stability diagram for the QQD device is shown in Fig. 4(b) of the main article.

| $C_{i=S,D}$ | $C_{M12}$ | $C_{M13}$ | $C_{G3\text{-}QD1}$ | $C_{M23}$ | $C_{M24}$ | $C_{G3\text{-}QD2}$ |
|---|---|---|---|---|---|---|
| 8.00 | 0.75 | 0.02 | 4.10 | 0.80 | 0.02 | 1.77 |
| $C_{G4\text{-}QD2}$ | $C_{G7\text{-}QD2}$ | $C_{M34}$ | $C_{G3\text{-}QD3}$ | $C_{G6\text{-}QD3}$ | $C_{G7\text{-}QD3}$ | $C_{G7\text{-}QD4}$ |
| 8.70 | 0.15 | 1.20 | 0.30 | 6.91 | 4.36 | 2.26 |

Based on the capacitance network model and the extracted capacitances listed in Table SI, we also calculate the charge stability diagram of the QQD in the resonance region as shown in Fig. 5(d) of the main article with $V_{G4} = -0.386$ V and $V_{G6} = -1.0535$ V. The results are shown in Fig. S4. Compared to the experimental results shown in Fig. 5(d) of the main article, the spacing between the two current lines corresponding to resonant transport through QD1 and QD2 and the spacing between the current lines corresponding to resonant transport through QD3 and QD4 found in the simulated charge stability diagram are again in good agreement with the experiment. However, at the avoiding crossings of different current lines, the simulated line shapes appear to be clearly different from the experiment ones. These differences are due to the fact that the quantum tunnel coupling mechanism has been neglected in our model simulations.[4,6]

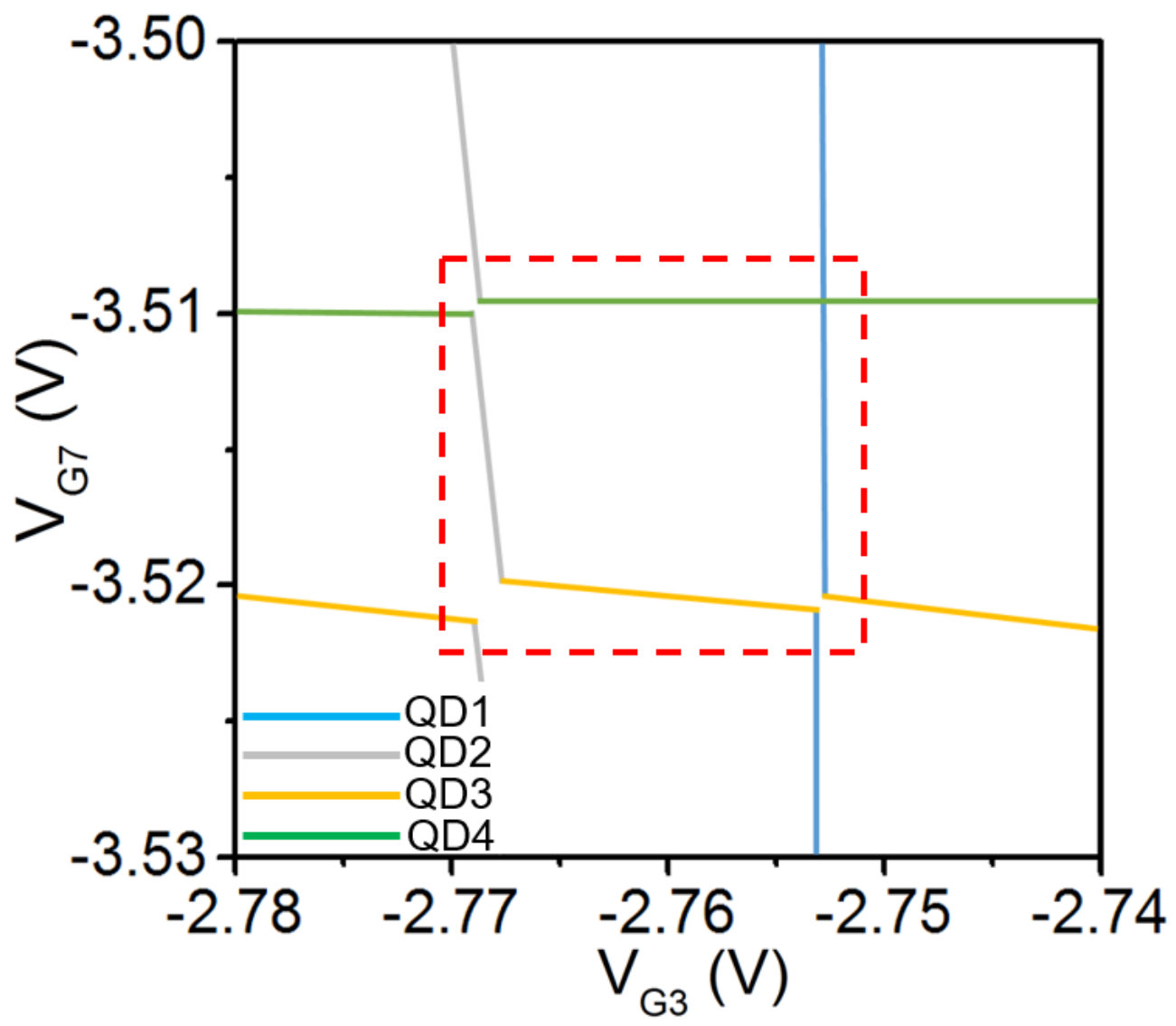

**Fig. S4** Calculated charge stability diagram for the nanowire QQD in the resonance region as shown in Fig. 5(d) of the main article with the voltages applied to plunger gates G4 and G6 set at $V_{G4} = -0.386$ V and $V_{G6} = -1.0535$ V, with the values of the capacitances given in Table SI. The blue, grey, yellow, and green lines represent the resonant current lines arising from tunneling via the energy levels of QD1, QD2, QD3, and QD4, respectively.

## V. Achieving the resonance between the energy levels of QD3 and QD4 in the QQD device studied in the main article

Figures S5(a) to S5(i) show the source-drain current $I_{ds}$ measured for the QQD device as a function of $V_{G3}$ and $V_{G7}$ (charge stability diagrams) at different voltage values of $V_{G6}$. Note that Figs. S5(a) and S5(e) are the same figures as Figs. 5(a) and 5(b) of the main article. The current lines marked with grey, yellow and green dash lines in Figs. S5(a) to S5(i) arise from resonant transport via the energy levels in QD2, QD3 and QD4, respectively. Figures S5(a) to S5(i) demonstrate that with decreasing $V_{G6}$, the spacing between a current line (labeled by 3) corresponding to resonant transport through the energy level of QD3 and a current line (labeled by 4) corresponding to resonant transport through the energy level of QD4 decreases, reaches a minimum at $V_{G6} = -1.0535$ V, and then increases. Thus, it is achieved that at $V_{G6} = -1.0535$ V, the energy levels of QD3 and QD4 are found to be on resonance with each other.

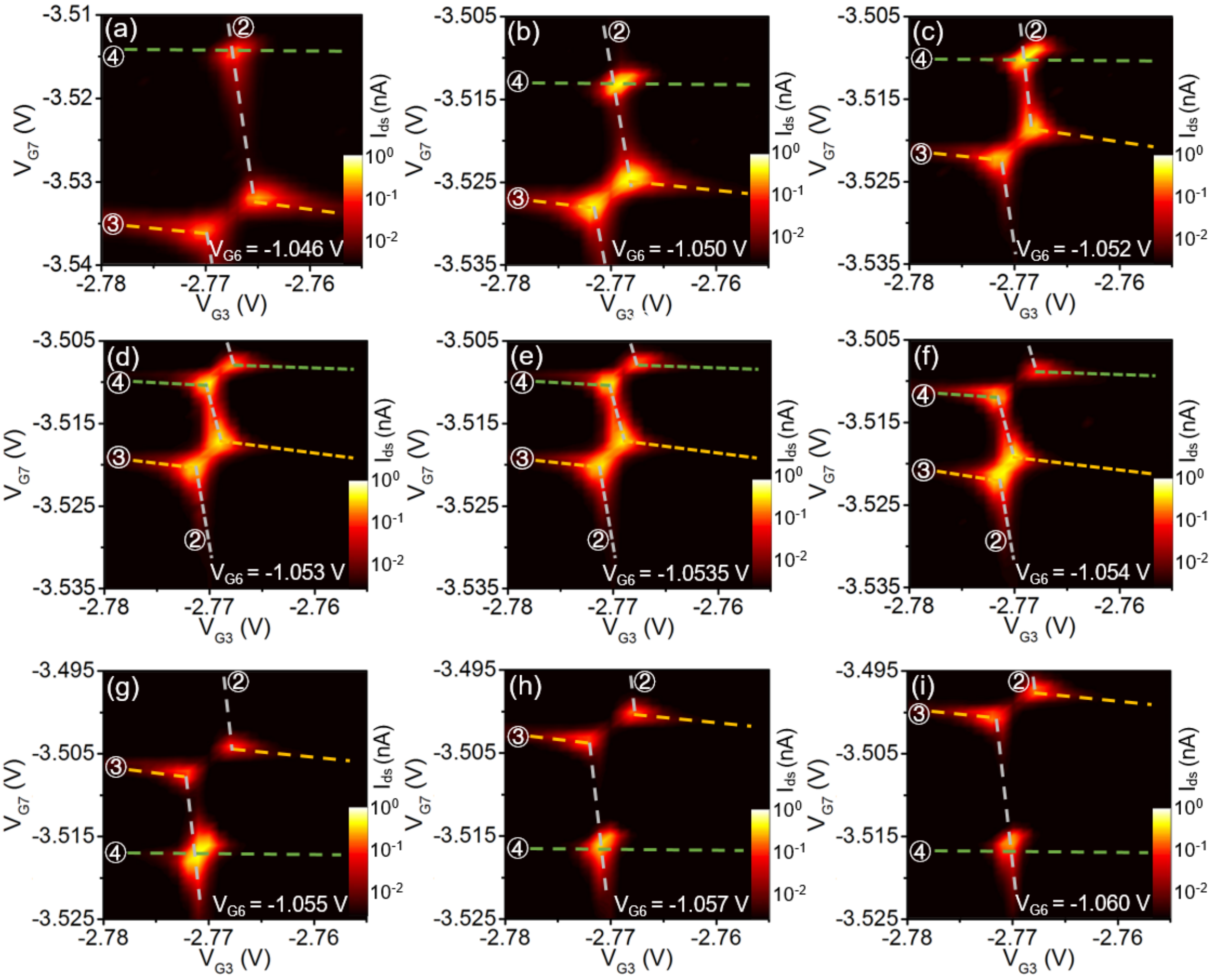

**Fig. S5** Charge stability diagrams measured for the QQD device at (a) $V_{G6} = -1.046$ V, (b) $V_{G6} = -1.050$ V, (c) $V_{G6} = -1.052$ V, (d) $V_{G6} = -1.053$ V, (e) $V_{G6} = -1.0535$ V, (f) $V_{G6} = -1.054$ V, (g) $V_{G6} = -1.055$ V, (h) $V_{G6} = -1.057$ V, and (i) $V_{G6} = -1.060$ V. The measurements are performed with the barrier gate voltages set at $V_{G1} = -2.55$ V, $V_{G5} = -3.35$ V and $V_{G9} = -3.855$ V, and the plunger gate voltages set at $V_{G4} = -0.384$ V and $V_{G8} = -0.94$ V. The source-drain voltage is set at $V_{ds} = 0.2$ mV in (a) and (b), and at $V_{ds} = 40$ μV in the other seven panels. The current lines marked by the grey, yellow and green dashed lines arise from resonant transport through QD2, QD3 and QD4, respectively. These current lines are also labeled by the corresponding QD numbers 2, 3 and 4.